\documentclass[longauth]{aa}

\usepackage{amssymb}
\usepackage{amsmath}
\usepackage{graphicx}
\DeclareGraphicsExtensions{.eps, .ps, .eps .gz, ps.gz}
\usepackage[]{hyperref}
\usepackage{color}

\bibpunct{(}{)}{;}{a}{}{,}

\newcommand{\kms}{$\,{\rm km\, s^{-1}}$~}

\begin{document}

\title{The VIMOS Public Extragalactic Redshift Survey (VIPERS)}
\subtitle{PCA-based automatic cleaning and reconstruction of survey spectra\thanks{based on
observations collected at the European Southern Observatory, Cerro Paranal, Chile, using the Very Large Telescope under programs 182.A-0886 and partly 070.A-9007.
Also based on observations obtained with MegaPrime/MegaCam, a joint project of CFHT and CEA/DAPNIA, at the Canada-France-Hawaii Telescope (CFHT), which is operated by the National Research Council (NRC) of Canada, the Institut National des Sciences de l’Univers of the Centre National de la Recherche Scientifique (CNRS) of France, and the University of Hawaii. This work is based in part on data products produced at TERAPIX and the Canadian Astronomy Data Centre as part of the Canada-France-Hawaii Telescope Legacy Survey, a collaborative project of NRC and CNRS. The VIPERS web site is  \url{http://www.vipers.inaf.it/}.}
}
\titlerunning{PCA cleaning of VIPERS spectra}
\author{
A.~Marchetti\inst{\ref{iasf-mi}}   
\and B.~Garilli\inst{\ref{iasf-mi}}          
\and B.~R.~Granett\inst{\ref{brera},\ref{unimi}}
\and L.~Guzzo\inst{\ref{brera},\ref{unimi}}      
\and A.~Iovino\inst{\ref{brera}}
\and M.~Scodeggio\inst{\ref{iasf-mi}}       
\and M.~Bolzonella\inst{\ref{oabo}}      
\and S.~de la Torre\inst{\ref{lam}}       
%
%GROUP A2:
% 
\and U.~Abbas\inst{\ref{oa-to}}
\and C.~Adami\inst{\ref{lam}}
\and D.~Bottini\inst{\ref{iasf-mi}}
\and A.~Cappi\inst{\ref{oabo},\ref{nice}}
\and O.~Cucciati\inst{\ref{unibo},\ref{oabo}}           
\and I.~Davidzon\inst{\ref{lam},\ref{oabo}}   
\and P.~Franzetti\inst{\ref{iasf-mi}}   
\and A.~Fritz\inst{\ref{iasf-mi}}       
\and J.~Krywult\inst{\ref{kielce}}
\and V.~Le Brun\inst{\ref{lam}}
\and O.~Le F\`evre\inst{\ref{lam}}
\and D.~Maccagni\inst{\ref{iasf-mi}}
\and K.~Ma{\l}ek\inst{\ref{warsaw-nucl}}
\and F.~Marulli\inst{\ref{unibo},\ref{infn-bo},\ref{oabo}} 
\and M.~Polletta\inst{\ref{iasf-mi},\ref{marseille-uni},\ref{toulouse}}
\and A.~Pollo\inst{\ref{warsaw-nucl},\ref{krakow}}
\and L.A.M.~Tasca\inst{\ref{lam}}
\and R.~Tojeiro\inst{\ref{st-andrews}}  
\and D.~Vergani\inst{\ref{iasf-bo}}
\and A.~Zanichelli\inst{\ref{ira-bo}}
%
% GROUP B: 
%
\and S.~Arnouts\inst{\ref{lam},\ref{cfht}} 
\and J.~Bel\inst{\ref{cpt}}
\and E.~Branchini\inst{\ref{roma3},\ref{infn-roma3},\ref{oa-roma}}
\and J.~Coupon\inst{\ref{geneva}}
\and G.~De Lucia\inst{\ref{oats}}
\and O.~Ilbert\inst{\ref{lam}}
\and T.~Moutard\inst{\ref{halifax}\ref{lam}}  
\and L.~Moscardini\inst{\ref{unibo},\ref{infn-bo},\ref{oabo}}
\and G.~Zamorani\inst{\ref{oabo}}
}
\offprints{ A.~Marchetti \\ \email{alida@lambrate.inaf.it} }
\institute{
%A1:
\and INAF - Istituto di Astrofisica Spaziale e Fisica Cosmica Milano, via Bassini 15, 20133 Milano, Italy \label{iasf-mi}%3
INAF - Osservatorio Astronomico di Brera, Via Brera 28, 20122 Milano
--  via E. Bianchi 46, 23807 Merate, Italy \label{brera}%1
\and  Universit\`{a} degli Studi di Milano, via G. Celoria 16, 20133 Milano, Italy \label{unimi}%2
\and INAF - Osservatorio Astronomico di Bologna, via Ranzani 1, I-40127, Bologna, Italy \label{oabo} %4
\and Aix Marseille Univ, CNRS, LAM, Laboratoire d'Astrophysique de
Marseille, Marseille, France  \label{lam}%5
%A2:
\and INAF - Osservatorio Astrofisico di Torino, 10025 Pino Torinese, Italy \label{oa-to}%5
\and Laboratoire Lagrange, UMR7293, Universit\'e de Nice Sophia Antipolis, CNRS, Observatoire de la C\^ote d’Azur, 06300 Nice, France \label{nice}%
\and Institute of Physics, Jan Kochanowski University, ul. Swietokrzyska 15, 25-406 Kielce, Poland \label{kielce}%15
\and National Centre for Nuclear Research, ul. Hoza 69, 00-681 Warszawa, Poland \label{warsaw-nucl}%23
\and Dipartimento di Fisica e Astronomia - Alma Mater Studiorum Universit\`{a} di Bologna, viale Berti Pichat 6/2, I-40127 Bologna, Italy \label{unibo}%17
\and INFN, Sezione di Bologna, viale Berti Pichat 6/2, I-40127 Bologna, Italy \label{infn-bo}%18
\and Aix-Marseille Université, Jardin du Pharo, 58 bd Charles Livon, F-13284 Marseille cedex 7, France \label{marseille-uni}
\and IRAP,  9 av. du colonel Roche, BP 44346, F-31028 Toulouse cedex 4, France \label{toulouse} 
\and Astronomical Observatory of the Jagiellonian University, Orla 171, 30-001 Cracow, Poland \label{krakow} %22
\and School of Physics and Astronomy, University of St Andrews, St Andrews KY16 9SS, UK \label{st-andrews}%11
\and INAF - Istituto di Astrofisica Spaziale e Fisica Cosmica Bologna, via Gobetti 101, I-40129 Bologna, Italy \label{iasf-bo}%25
\and INAF - Istituto di Radioastronomia, via Gobetti 101, I-40129,
Bologna, Italy \label{ira-bo}%26
%B:
\and Canada-France-Hawaii Telescope, 65--1238 Mamalahoa Highway, Kamuela, HI 96743, USA \label{cfht}%6
\and Aix Marseille Univ, Univ Toulon, CNRS, CPT, Marseille, France \label{cpt}%7
\and Dipartimento di Matematica e Fisica, Universit\`{a} degli Studi Roma Tre, via della Vasca Navale 84, 00146 Roma, Italy\label{roma3} %10
\and INFN, Sezione di Roma Tre, via della Vasca Navale 84, I-00146 Roma, Italy \label{infn-roma3}%28
\and INAF - Osservatorio Astronomico di Roma, via Frascati 33, I-00040 Monte Porzio Catone (RM), Italy \label{oa-roma}%29
\and Department of Astronomy, University of Geneva, ch. d’Ecogia 16, 1290 Versoix, Switzerland \label{geneva}%12
\and INAF - Osservatorio Astronomico di Trieste, via G. B. Tiepolo 11, 34143 Trieste, Italy \label{oats}%13
%C:
\and Department of Astronomy \& Physics, Saint Mary's University, 923 Robie Street, Halifax, Nova Scotia, B3H 3C3, Canada \label{halifax}%13
\and Institute of Cosmology and Gravitation, Dennis Sciama Building, University of Portsmouth, Burnaby Road, Portsmouth, PO13FX \label{icg}%32
\and Institut Universitaire de France \label{inst-france}%30
\and Institute d'Astrophysique de Paris, UMR7095 CNRS, Universit\'{e} Pierre et Marie Curie, 98 bis Boulevard Arago, 75014 Paris, France \label{iap} %19
\and Institute for Astronomy, University of Edinburgh, Royal
Observatory, Blackford Hill, Edinburgh EH9 3HJ, UK \label{roe}%14
%
%EXTRA:
%\and Center for Theoretical Physics, Al. Lotnikow 32/46, 02-668 Warsaw, Poland \label{warsaw-theory} %31 Siudek
%\and Department of Particle and Astrophysical Science, Nagoya University, Furo-cho, Chikusa-ku, 464-8602 Nagoya, Japan \label{nagoya}%16
%\and Max-Planck-Institut f\"{u}r Extraterrestrische Physik, D-84571 Garching b. M\"{u}nchen, Germany \label{mpe}%20
%\and Astronomical Observatory of the Jagiellonian University, Orla 171, 30-001 Cracow, Poland \label{cracow}%22
%\and Universit\"{a}tssternwarte M\"{u}nchen, Ludwig-Maximillians Universit\"{a}t, Scheinerstr. 1, D-81679 M\"{u}nchen, Germany \label{munich} %24
%\and Dipartimento di Fisica, Universit\`a di Milano-Bicocca, P.zza della Scienza 3, I-20126 Milano, Italy \label{bicocca}%  for Pezzotta
%\and Universit\'{e} de Lyon, F-69003 Lyon, France \label{lyon}%8
%
}

\date{Received ; accepted}

\abstract
{Identifying spurious reduction artefacts in galaxy spectra is a challenge for large surveys. }
{We present an algorithm for identifying and repairing residual spurious features in sky-subtracted galaxy spectra with application to the VIPERS survey.}
{The algorithm uses principal component analysis (PCA) applied to the galaxy spectra in the observed frame to identify sky line residuals imprinted at characteristic wavelengths.  We further model the galaxy spectra in the rest-frame using PCA to estimate the most probable continuum in the corrupted spectral regions, which are then repaired.}
{We apply the method to 90,000 spectra from the VIPERS survey and compare the results with a subset where careful editing was performed by hand. We find that the automatic technique does an extremely good job in reproducing the time-consuming manual cleaning and does it in a uniform and objective manner across a large data sample.  The mask data products produced in this work are released together with the VIPERS second public data release (PDR-2).}
{}

\titlerunning{VIPERS: automatic cleaning of spectra}
\authorrunning{Marchetti et al.}
\maketitle

\section{Introduction}
Large surveys of galaxy redshifts represent one of the primary means to 
explore the structure of the Universe and the evolution of galaxies. These include wide-angle 
surveys at relatively low redshift, notably the SDSS \citep{york00} and 2dFGRS 
\citep{colless01} and deeper, narrower probes including VVDS \citep{lefevre05,garilli08}, 
DEEP2  \citep{newman13} and zCOSMOS \citep{lilly09} [see \citet{guzzo14} for a more complete review of current and past surveys].  The recently completed
VIMOS Public Extragalactic Redshift Survey (VIPERS) has built a sample 
that is at the same time deep, densely sampled and covers a volume similar to the 2dFGRS, 
but at $z= [0.5,1.2]$ \citep{scodeggio16,guzzo14,garilli14}.  

These surveys provide unique insight into both cosmology and galaxy formation.  The spectra of extragalactic sources can enlighten our understanding of the underlying processes of galaxy evolution through the measurements of physical properties such as star formation, metallicity, gas content and rotational velocity.   The observed redshift contains not only information on the galaxy distance from the uniform Hubble expansion, but also the imprint of peculiar motions produced by the growth of structure. Statistically, this information can be extracted and used to test the nature of gravity
\citep[see e.g. the parallel papers by][]{delatorre16,hawken16,pezzotta16,wilson16}. Knowledge of precise distances to galaxies also enables us to map the cosmic web and characterize galaxies with respect to the local density field; this is the starting point
to study the interplay between galaxy properties and their environment \citep[see the parallel paper by][]{cucciati16}. 

Spectra obtained from ground--based optical surveys
suffer from contamination from signal coming from the Earth's
atmosphere (besides instrumental and data reduction artifacts). This
sky emission affects the identification and measurements of emission and absorption features, 
which are used to determine the redshift; they will also corrupt estimates of line intensities, through which galaxy properties are characterised.
These defects can be cured manually, when the number of spectra is small. In large modern surveys, however, automatic data reduction pipelines have become mandatory, as to efficiently manage the large quantity of data 
\citep[e.g.][and references therein]{stoughton02,garilli10}.

Spectroscopic data reduction pipelines perform the subtraction of sky lines and other known features; however, this operation is not free of errors, depending on various effects such as instrument optical distortions or the presence of fringing.  As a result, after automatic sky subtraction, spurious residuals may be left and contaminate the spectrum, especially at $\lambda>$ 7500\smash{\AA}, where the sky emission becomes more and more dominant. For this reason, the processed spectra are usually inspected and often cleaned by human intervention, e.g. 
substituting the corrupted portion with a sensible interpolation. This cleaning procedure facilitates and improves the quality of any spectral measurement performed on the calibrated spectrum (from the redshift to line intensities and spectral indices). 

Unfortunately, repairing these defects is not always
straightforward, and in some cases may require the investment of an
important amount of time.  This is not feasible for the large numbers of spectra implied by the modern industry of redshift surveys, with hundreds of thousands to millions of spectra collected within the same project.  Additionally, such cleaning would be intrinsically subjective: different operators will apply different styles of cleaning  across the same data, introducing some inhomogeneity.  The only appropriate way to perform an efficient and objective cleaning of redshift survey spectra from sky residuals or similar features is then to implement a fully-controlled, automatic procedure.  
  
In this paper we describe the automatic pipeline we have developed within the VIPERS project. 
The algorithm identifies the position of residual artefacts
that appear in the sky-subtracted spectra (\emph{observed
  spectra} from now onwards) and create a mask that matches their position; whenever
possible, the corrupted spectral section is reconstructed and repaired. Both the identification and reparation of the affected spectral sections are based on the application of Principal Component Analysis \citep[ PCA][]{karhunen47,connolly95,yip04}. 

A PCA-based sky subtraction was adopted by \citet{wild05} for the SDSS spectra: a set of sky emission templates was built by computing the principal components of the sky spectra, observed with a number of dedicated fibers. The best-fitting sky contribution to each galaxy spectrum was then estimated and subtracted.  Such an approach is appropriate for modeling and subtracting the sky spectra obtained with a fibre--fed spectrograph, but with VIPERS we face a different issue. VIMOS is a slitlet multi-object spectrograph, in which the sky is extracted from the fraction of slit adjacent to the object and then subtracted. This is done automatically by the data reduction pipeline \citep{scodeggio05}. All works well when the adjacent rows of sky spectrum are all aligned in wavelength with those containing the object spectrum.  Unfortunately, optical distortions and fringing on the CCD surface break this symmetry: sky emission lines are distorted along the slit, thus leading to a sub-optimal subtraction that leaves residual features on the processed spectrum. 
Being related to the brightest sky features, these residuals appear at characteristic wavelengths.  

The idea we have successfully developed in this paper has been to identify these residuals through a template spectrum, which is obtained by applying the PCA to the {\sl observed-frame} galaxy spectra. In the observed frame, in fact, sky artifacts will sum up, while galaxy features will be, to some extent, suppressed as they appear at different redshifts. 
Given the stochastic nature of the residuals, however, this technique cannot be expected to reproduce the exact intensity and profile of each feature. Significant information will be contained in the high-order eigenvectors of the PCA (where less and less common details are encoded); these will get more and more mixed with real features from the galaxy spectra (since many objects are at similar redshift, thus sharing similar spectral structures  in the observed frame).  

We therefore have to limit ourselves to the first few eigenvectors (or {\sl eigenspectra}, as we call them), where the long redshift baseline of the survey guarantees that the main galaxy features are practically washed out. 
Under these conditions, the PCA reconstruction will not exactly reproduce the shape and intensity of the sky residuals, but can still be used to define a {\sl mask} that marks the corrupted spectral ranges. 

After the determination of the spectral sections to be masked we compute a realistic model of the spectrum to reconstruct the affected regions.  This further step, unlike the masking procedure, requires the knowledge of galaxy redshifts.  The contaminated regions we find are reconstructed by a second application of the PCA, this time performed in the galaxy {\sl rest frame} \citep{marchetti13}.  

Although designed for and calibrated on the VIMOS low-resolution spectra \citep[see][for details]{scodeggio16}, the procedure is quite general and can be easily transported to other surveys.
The paper is structured as follows: in \S 2  we briefly introduce the data on which the method has been developed (VIPERS spectra), in \S 3 we make a general overview of the PCA method, in \S 4 we describe the residuals masking procedure, in \S 5 its application to spectra, in \S 6 we describe the repairing of spectra within the masked regions and its advantages. In \S 7 we summarize and draw the conclusions. In the Appendix  we compare the automatic masking of VIPERS spectra to the manual one and discuss the results.

\section{The VIPERS survey}

The VIPERS survey spans an overall area of $ 23.5$ deg$^2$ over the W1 and W4 fields of the 
Canada-France-Hawaii Telescope Legacy Survey Wide (CFHTLS-Wide).  The VIMOS
multi-object spectrograph \citep{lefevre03} was used to cover these two regions
through a mosaic of 288 pointings, 192 in W1 and 96 in
W4.  
Galaxies were selected from the CFHTLS catalogue to a limit of
$i_{AB}<22.5$, applying an additional 
$(r-i)$ vs $(u-g)$ colour pre-selection that efficiently and robustly removes
galaxies at $z<0.5$.  Coupled with a highly optimised observing strategy
\citep{scodeggio09}, this doubles the mean galaxy sampling efficiency in the redshift range of interest, compared to a purely magnitude-limited sample, bringing it to 47\%. 

Spectra were collected at moderate
resolution using the LR Red grism (7.14~\smash{\AA}/pixel, corresponding to an average $R\simeq 220$), providing a wavelength coverage of 5500-9500~\smash{\AA}. The typical redshift error 
for the sample of reliable redshifts (see below for definitions) is
$\sigma_z=0.00054(1+z)$; this  corresponds to an error on a galaxy peculiar velocity at any redshift of 163~\kms.  

The data were processed with
the {\sc Pandora Easylife} \citep{garilli10} reduction pipeline. Redshifts and quality flags are measured with the {\sc Pandora EZ}
(Easy Z) package \citep{garilli10}.  
The redshift and flag assigned by the {\sc Pandora} pipeline were visually checked and validated, typically by two team members for each spectrum.  The quality flag, in the form $\pm XY.Z$, 
indicates the confidence level of the redshift measurement. The most relevant digit is the second, $Y$, which can take  values 0, 1, 2, 3, 4, in order of increasing confidence; the value 9 is reserved for measurements based on a single emission line.
Measurements with $Y = 2$ or larger define what the sample of reliable redshifts, which is used for statistical investigations, guaranteeing a confidence of 96.1\% \citep[see][for a description of the quality flag
scheme]{scodeggio16}. Flag 1 redshifts are 
highly uncertain at the 50\% confidence level. Flag 0 objects are not part of the officially released PDR-2 catalogue, as they correspond to spectra for which a redshift could not be assigned. This are included in the analysis presented here, since they often provide the most information about artefacts from sky residuals. Thus, this work uses the whole set of 97,414 observed spectra, whose detail is given in Table 2 of \citet{scodeggio16}.

Together with object spectra, the VIPERS survey database provides the sky spectra and the noise spectra associated to every observed spectrum \citep{garilli10}.  The data from the final VIPERS Public Data Release (PDR-2) are available at \url{http://vipers.inaf.it}.

\section{PCA on spectra}

Principal component analysis (PCA) is a non-parametric way to reduce the complexity of high-dimensional datasets while preserving
the majority of the information.  This is possible when strong correlations exist in the data as is the case for galaxy spectra.
Galaxy spectra share many common features but yet are unique.

PCA consists of a linear transformation that changes the frame of reference from the observed, or natural, one to a frame of reference that
highlights the structure and correlations in the data.  The transformation aligns the principal axes with the directions of maximum variance in the data and is computed by diagonalising the data correlation (or covariance) matrix.  When applied to spectra with flux measurements $f_{\lambda}$ in $M$ bins the correlation matrix is given by

\begin{equation}
C_{\lambda_1,\lambda_2}=\frac{1}{N-1}\sum_{i=1}^N f_{\lambda_1}^{i}f_{\lambda _2}^{i}\label{eq:corr},
\end{equation}
where $i$ indexes the $N$ spectra in the sample and $\lambda_1$ and
$\lambda_2$ index wavelength bins of the $M^2$ element correlation matrix.

The eigenvectors $e^i_{\lambda_j}$ of the sample, obtained diagonalizing the  correlation matrix

\begin{equation}
C_{\lambda_1,\lambda_2}=\sum_{i=1}^M e^i_{\lambda_1}\Lambda_i e^i_{\lambda_2}.
\end{equation}
represent the axes of the new coordinate system.
The basis one obtains will be made up by
orthogonal (i.e. uncorrelated) eigenvectors which
are linear combinations of the original variables.  
The eigenvalues give the variance of the data in the orthogonal space and
may be used to order the eigenvectors.  By using only the most significant
eigenvectors corresponding to the largest eigenvalues we can reconstruct
most of the statistical information in the dataset.  

Our data consist of $N$ galaxy spectra each with $M$ wavelength bins. Since the eigenspectra have the shape of spectra we address them as 
 \emph{eigenspectra}.  When the spectra are kept in the observed frame, the signature of sky residuals is a coherent feature while the signal from astrophysical emission/absorption features present in our targets cancels out due to the wide range of different redshifts. Nonetheless, we will still find a smooth signal representing the superposition of all spectral continua, 
even if they are shifted with respect to each other according to their redshift. To eliminate this, we subtract the continua before computing the correlation matrix.

Each of the observed-frame spectral energy distributions, namely $f_{\lambda}$, can be expressed as a
sum of the $M$ eigenvectors with a set of $M$ linear coefficients.  
We will truncate the sum to use only the first $K\ll M$ components:
\begin{equation}
\hat{f}_{\lambda}=\sum_{i=1}^K a_i e^i_{\lambda}\label{eq:pca},
\end{equation}
where $a_i$ are the linear coefficients.  Projecting the observed-frame spectra onto the leading eigenspectra will give our best estimate of the locations and strengths of the sky residuals.  We refer to the projection,  $\hat{f}_{\lambda}$ in Eq. \ref{eq:pca}, as the \emph{sky residuals spectrum}.  The choice of the number of components to take may be made based upon the relative power of each:
\begin{eqnarray}
P(e_i)=\frac{\Lambda_i}{\Sigma_{k}^{tot}\Lambda_k}
\end{eqnarray}
where $\Lambda_i$ stands for the $i$-th eigenvalue, related to the $i$-th eigenspectrum $e_i$.

\section{The sky residuals eigenspectra}
The first step of the method is to obtain the sky residuals eigenspectra.  This is performed after subtracting the continuum and normalizing the spectra by the scalar product.  We estimate the continuum by convolving with a Gaussian kernel with width $\sigma=50$ pixels, corresponding to 355\AA.  After computing the eigenspectra, they are ordered with decreasing eigenvalue, such that the most common features within the spectra are contained in the first few eigenspectra. 

\begin{figure}
\includegraphics[scale=1.2]{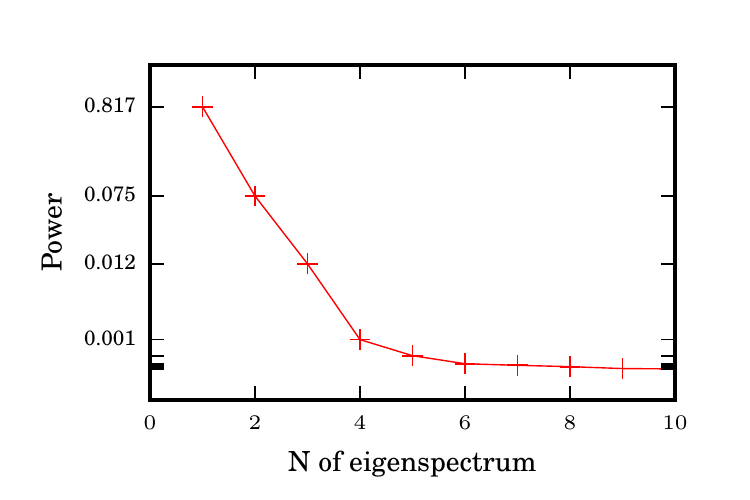}
\caption{\small{The power associated with the first 10 eigenspectra.  The labels on the vertical axis indicate the abscissae of the data points. }}\label{power}
\end{figure}
\begin{figure}

\center\includegraphics[scale=1.2]{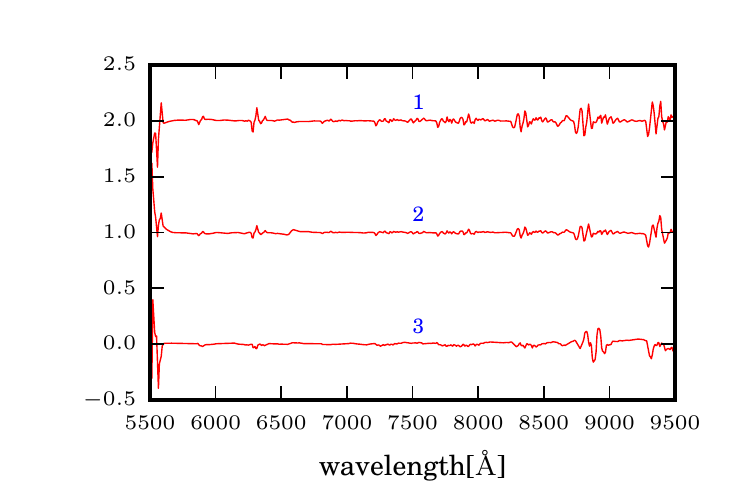}
\caption{\small{The 3 principal components from the observed-frame VIPERS dataset. The first and second eigenspectra have been offset by 2 and 1 respectively, for visualization convenience.
}}\label{skypre}
\end{figure}
\begin{figure}
\center\includegraphics[scale=1.2]{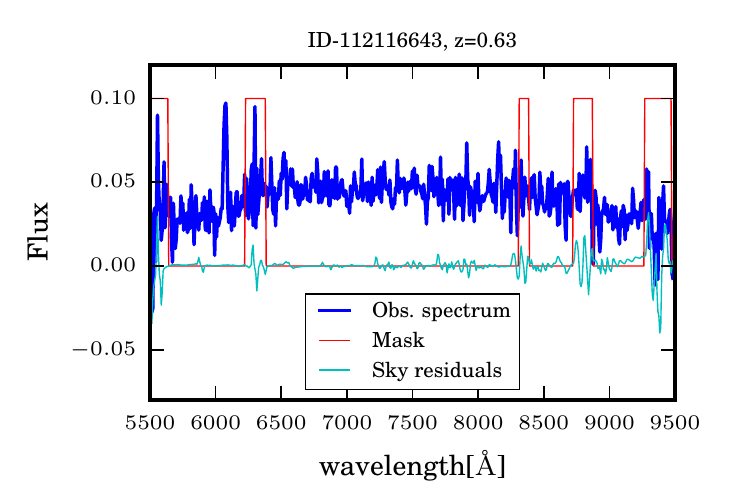}
\caption{\small{An example of sky residuals spectrum (thin cyan) and its relevant observed spectrum (thick blue), from the VIPERS survey. The red straight lines indicate where the mask is applied, on the basis of the sky residuals spectrum.
}}\label{rec}
\end{figure}

To determine how many components to keep, we consider the size of the corresponding eigenvalues.
Fig.~\ref{power} shows the power associated to the first 10 eigenspectra.
The first eigenspectrum gives the average contribution of the sky residuals and it alone explains nearly 82\% of the variance in the dataset.
The second and third components encode 7.5\% and 1.2\% of the remaining information giving a total of 90\% in the first three components.
The information in the fourth one is significantly lower at 0.16\% and the value of each higher order eigenspectrum decreases steadily. 
On the basis of this distribution of power as a function of the number of eigenspectra, we decided to use the first three to characterize the residual spectra.  Using more does not improve the results and can lead to over-fitting astrophysical features.
We thus construct a basis from the three most significant eigenspectra, as shown in Fig.~\ref{skypre}.
The continuum-subtracted spectra are projected onto this basis to compute the residual spectra (Fig.~\ref{rec}).

The main aim of this analysis is to define a mask for each
spectrum, in correspondence with the more intense sky residual
features of the associated reconstructed sky residuals spectra. While the position of the sky residual features is recovered with reasonable accuracy, their intensity is often slightly over- or under- estimated: this is a consequence of using only few eigenspectra (see \citealt{marchetti13}). However, as the aim is to determine the position of the sky residuals, other than to capture their precise strength, these discrepancies in intensity are not important. 

\section{Automatic masking  of the spectra}
 
\subsection{Identifying the location of common sky features}
We estimate the residuals of contamination for each galaxy spectrum, the \emph{sky residuals spectrum}, by projecting the spectra onto the basis of eigenspectra. The sky residuals spectrum is used to determine the threshold for masking.  For each sky residuals spectrum, we compute the mean value and standard deviation $\sigma$.  
We mask all wavelengths where the sky residuals spectrum exceeds the mean of the spectrum by greater than $k\sigma$. The parameter $k$ is set according to the characteristics of the dataset. For VIPERS, we adopted 1.2$\sigma$
at wavelengths shorter than 7500\AA, and 1.8$\sigma$ at wavelengths longer than 7500\AA, due to the higher contamination. These thresholds have been chosen empirically: the values were set to select the known sky lines in a subset of representative spectra.

\begin{figure}
\includegraphics[scale=1.2]{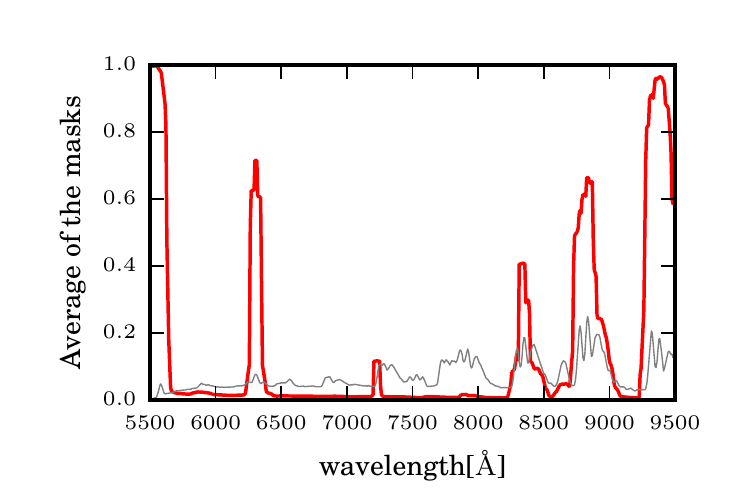}
\caption{\small{Average of the masks for the VIPERS sample (thick red); a VIPERS sky spectrum is depicted in grey (thin).}}\label{summ}
\end{figure}

\begin{figure}
\includegraphics[scale=1.2]{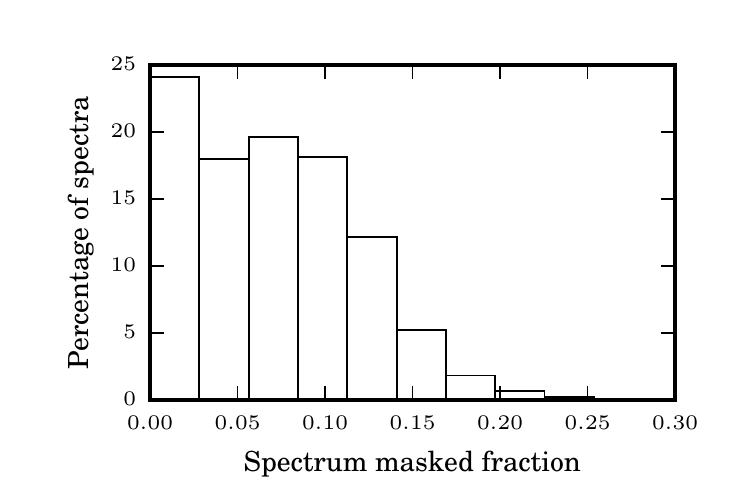}
\caption{\small{Histogram for the distribution of the fraction of masked spectrum after the sky/zero-order residuals masking procedure (the regions at the edges of the spectra have been excluded).}}\label{mfrac}
\end{figure}

Fig.~\ref{summ} shows the frequency of masked wavelengths after applying the algorithm to the VIPERS dataset.  
Around 65\% of spectra have been masked in correspondence of the $\lambda$=6300\smash{\AA} sky line and of the OH group at $\sim\lambda$=8700\AA; about 40\% have been masked around $\lambda$=8300\AA, and only the 10\% at $\sim\lambda$=7300\AA. Nearly all spectra are masked at the upper and lower wavelength limits of the spectrum, where there is a significant contribution from fringing and calibration issues. In particular, at the lower limit there is the presence of the 5577\smash{\AA} sky line residual combined with the fall-off in the detector sensitivity.  The extent of the masking is summarised in Fig.~\ref{mfrac}.  For half of the spectra, the mask covers less than 5\% of the wavelength range.

\begin{figure}
\center\includegraphics[scale=1.2]{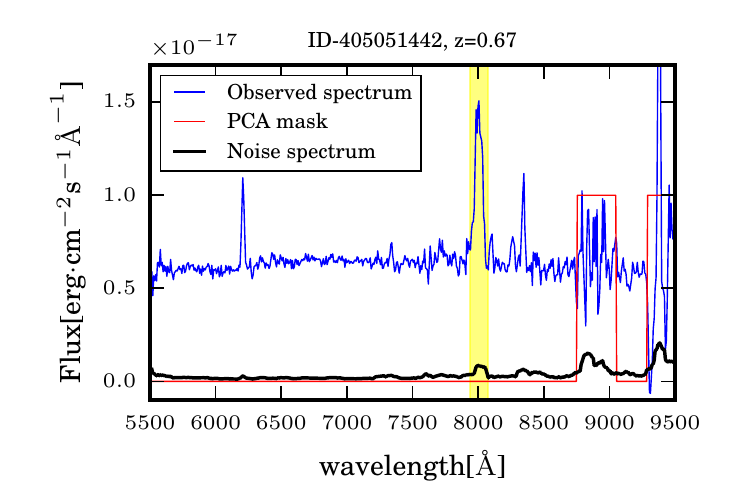}
\caption{\small{Example of zero-order residual in a VIPERS spectrum (top, highlighted by the yellow vertical bar); at the bottom we show the corresponding noise spectrum.  The clear excess corresponding to the zero-order position is used to supplement the sky residual mask (red line), as to account for this extra contribution in the final cleaning and repairing. 
}}\label{zero}
\end{figure}

Specifically for the VIPERS spectra, during this phase we also accounted for the extra residuals originating from the subtraction of the zero-order images of bright objects, as shown in the example of Fig.~\ref{zero}. These extra features were identified in the noise spectra delivered by the VIPERS reduction pipeline through a simple thresholding. Their position and size was thus added to the sky residual mask for subsequent treatment.

\begin{figure}
\includegraphics[scale=0.6]{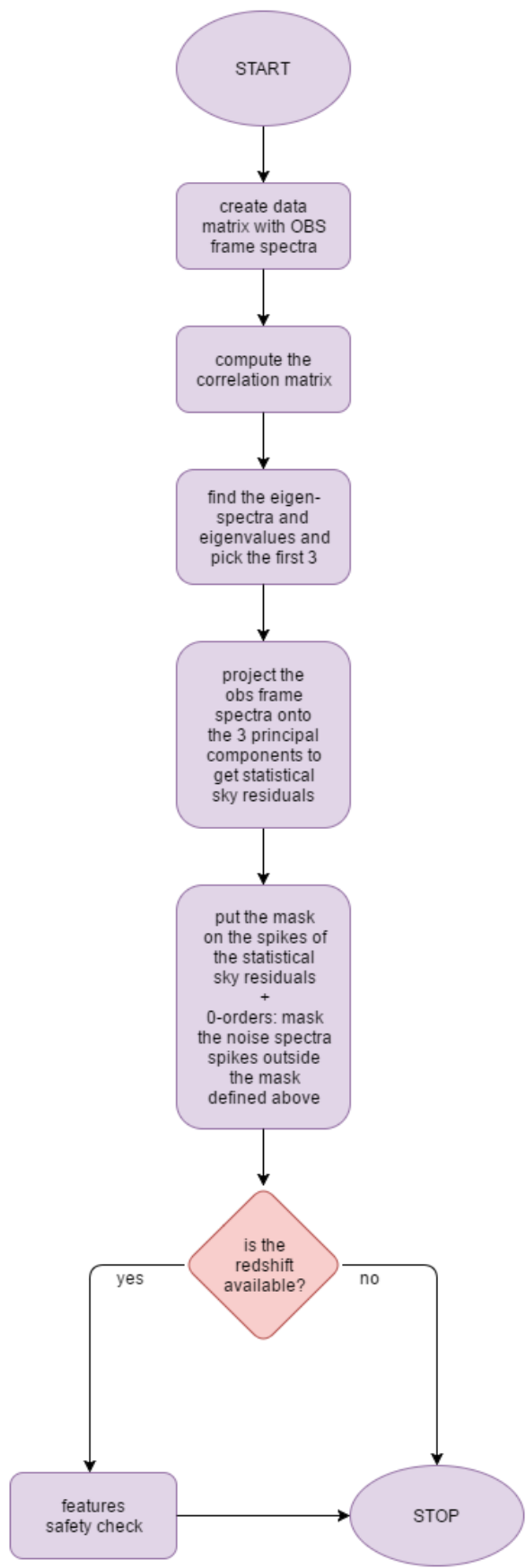}
\caption{\small{Scheme of the automatic masking procedure.}}\label{flow1} 
\end{figure}

The process described so far is summarised in the flow chart of Fig.~\ref{flow1}.

\section{Repairing the spectra}

\begin{figure}
\includegraphics[scale=1.55]{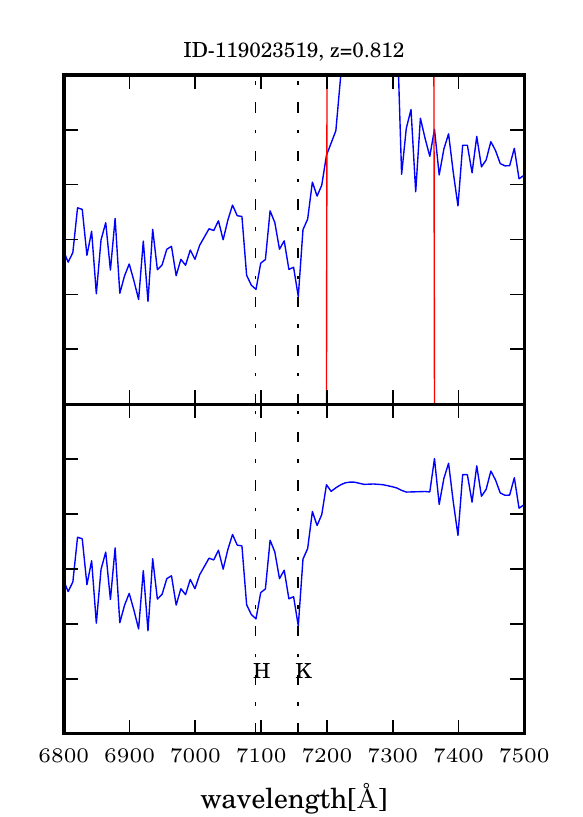}
\caption{\small{\emph{Top panel}. Zoom of a VIPERS spectrum with strong residual at the right of D4000\smash{\AA} break. The region of the mask is delimited by the red vertical lines. \emph{Bottom panel}. The same VIPERS spectrum after repairing the portion affected by the residual. 
}}\label{badline}
\end{figure}

We next compute an estimate for the galaxy continuum in masked regions allowing us to repair the contaminated data.  This is helpful not only for visual inspection of the spectra but aids the measurement of spectral features as well.  For example, line measurement tools require estimates of the continuum which may be unreliable due to spurious artefacts.  Fig.~\ref{badline}(\emph{top}) shows the D4000 break for a VIPERS spectrum affected by an artefact that prevents the proper measurement of the intensity of the break.   

\begin{figure}
\includegraphics[scale=1.2]{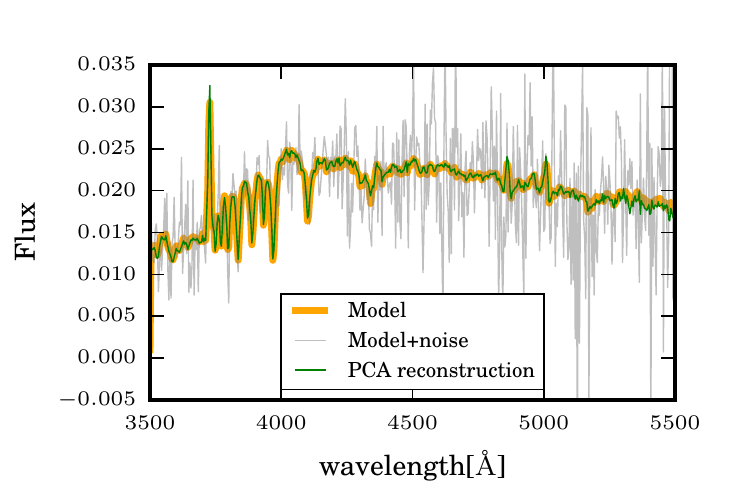}
\caption{\small{Comparison between a model spectrum (orange thick) and the PCA reconstruction of the same spectrum (thin green) after degrading it with noise (in soft grey the degraded spectrum).}}\label{test}
\end{figure}

Our approach to reconstruct and repair the spectrum is based on the PCA model in the rest-frame \citep{marchetti13}.   The shift to rest-frame can only be made if the redshift is known; thus, we apply the procedure only to galaxies with redshift quality flag $>=1$.  Futhermore, the wavelength range in the rest-frame is limited by the redshift range of the sample.  To have sufficient spectra we limit the sample to the redshift range to $0.4<z<1.4$.

 After shifting the spectra to the rest frame, we compute the eigenspectra as described in \citet{marchetti13}.  We use the most significant three eigenvectors to reconstruct the spectra continuum.   

The PCA with three components was found to accurately reconstruct the continuum of VIPERS spectra in \citet{marchetti13}.
Here we further demonstrate the accuracy of the reconstruction using mock spectra built from linear combinations of 
Bruzual-Charlot \citep{bruzual03} and Kinney-Calzetti \citep{kinney96} templates, also described in \citet{marchetti13}.  The mock spectra were redshifted and degraded with Gaussian noise to mimic the properties of the VIPERS spectra.  Using a sample of 20,000 mock spectra we estimated the first three eigenspectra.  We then project the mock spectra onto the basis of eigenspectra to compute the reconstruction.  We plot an example spectrum in Fig.~\ref{test} showing excellent agreement between the model and the reconstruction at all wavelengths.  While for the emission lines the discrepancies can be up to $\sim$25\% \citep{marchetti13}, the continua are always well reproduced; in particular from the test spectra, we found that the discrepancies between reconstructed and model continua are on average lower than $\sim$1.6\%, and are never worse than $\sim$5\%, where such a discrepancy is obtained in rare cases ($<0.1$\%).

\begin{figure}
\flushleft\includegraphics[scale=1.25]{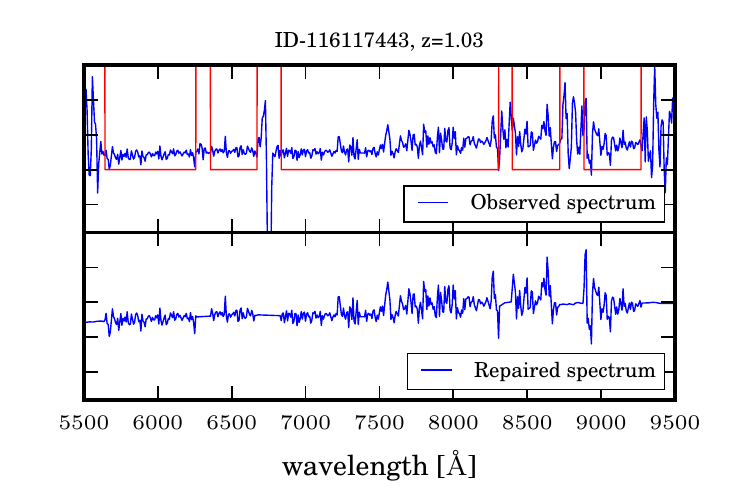}
\caption{\small{Example of a VIPERS spectrum (blue) and the corresponding mask (red straight lines) (top panel) and its PCA repairing in the masked portions (bottom).
}}\label{fill}
\end{figure}

Since the intensity of line features cannot be reproduced precisely we introduce a ``line safeguard'' and do not use the reconstruction at the locations of known features.  For VIPERS we ensure that the reconstruction does not substitute the most prominent emission lines
(e.g. [OII], H$\beta$, [OIIIa], [OIIIb], H$\alpha$ for galaxies), and the D4000 break. The safeguard was necessary for 30\% of VIPERS spectra for which a mask fell on a known feature.

An example of rest-frame repairing within the mask regions is shown in
Fig.~\ref{fill}.  After the repairing, the determination of the intensity
of the galaxy spectral features is easier and more reliable, as shown in 
Fig.~\ref{badline} \emph{bottom}.

\begin{figure}
\center\includegraphics[scale=0.57]{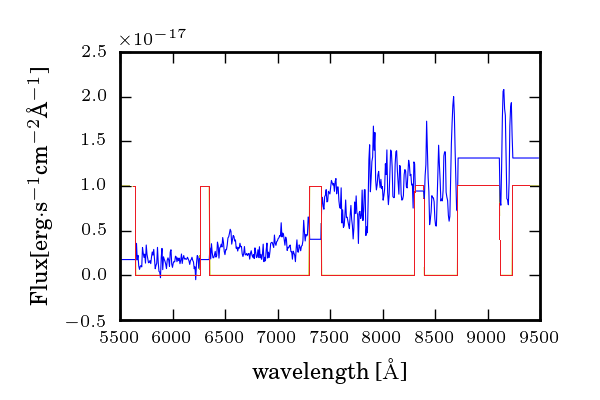}
\caption{\small{A VIPERS stellar spectrum repaired with points at constant value at the level of the continuum, computed near the regions of the mask (red).
}}\label{zzero}
\end{figure}

\begin{figure}
\includegraphics[scale=0.6]{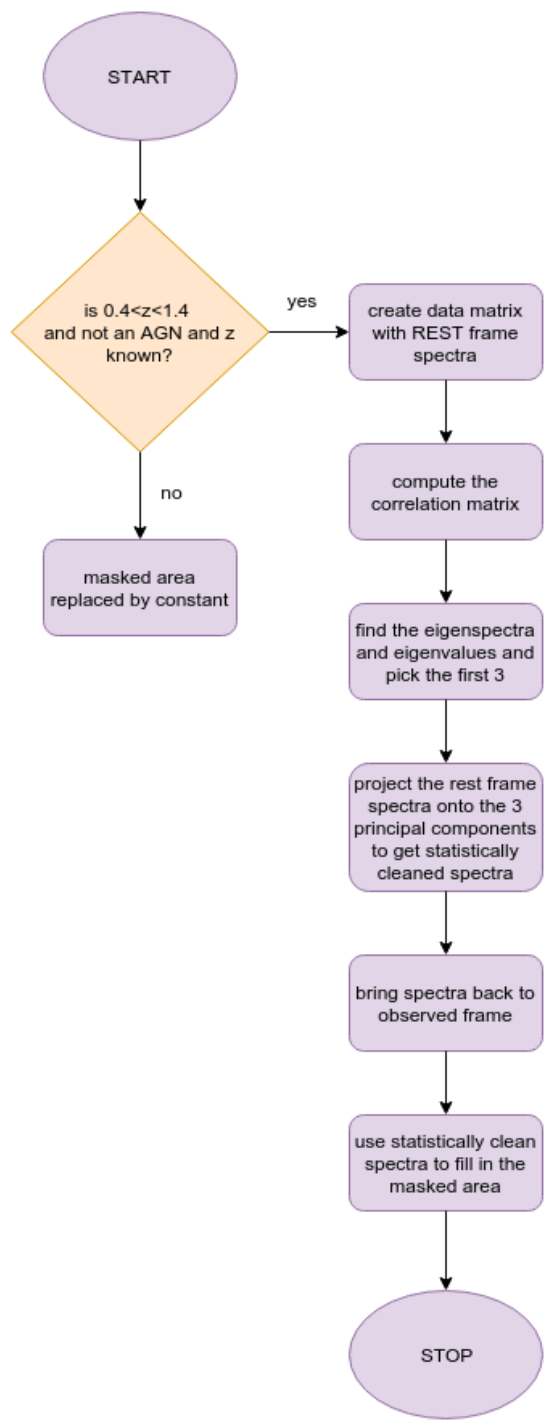}
\caption{\small{Scheme of the automatic repairing procedure}}\label{flow2}
\end{figure}

Not all spectra may be repaired using this procedure; sources outside the redshift range $0.4<z<1.2$ or sources without a measured redshift cannot be modeled using the PCA.  For these sources we use a constant interpolation to repair the masked regions, see Fig.~\ref{zzero}.  Additionally, for active galaxies (AGN), the PCA reconstruction based upon galaxy eigenspectra is not applicable. \citet{marchetti13} found that the eigenspectra computed from the full survey could not represent rare AGN spectra well.  Thus, for VIPERS spectra identified as AGN we have used constant interpolation to repair the masked regions.
The steps followed to create the automatic repairing described here are schematically listed in the flow chart of Fig.~\ref{flow2}.

\section{Comparison and combination of automatic and manual cleaning for the VIPERS data}

 For VIPERS spectra, a significant amount of time has been spent by the VIPERS team to manually clean the spectra from sky residuals or other artifacts, producing many careful manual edits. To check the reliability and efficiency of  the automatic procedure, we compared $\sim$ 500 automatically masked and repaired spectra with their corresponding manually edited spectra. 

\begin{figure*}
\center\includegraphics[scale=0.7]{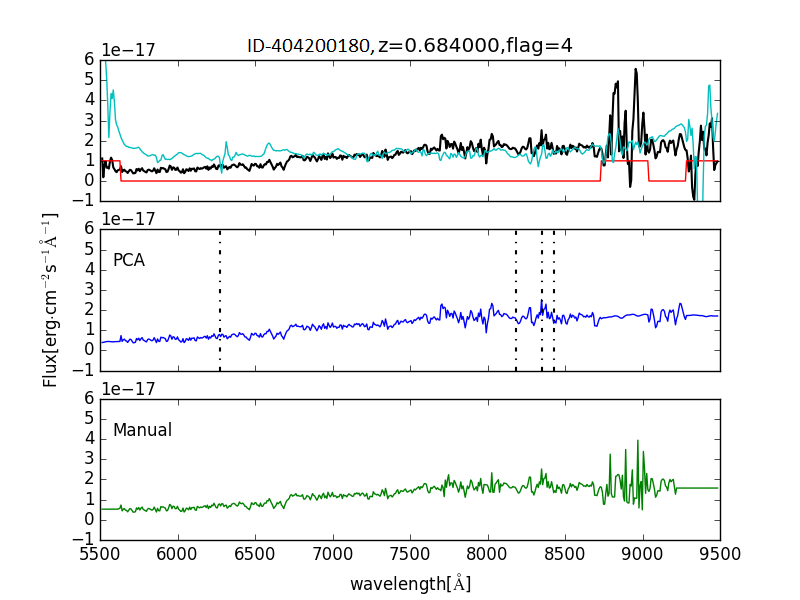}
\center\includegraphics[scale=0.7]{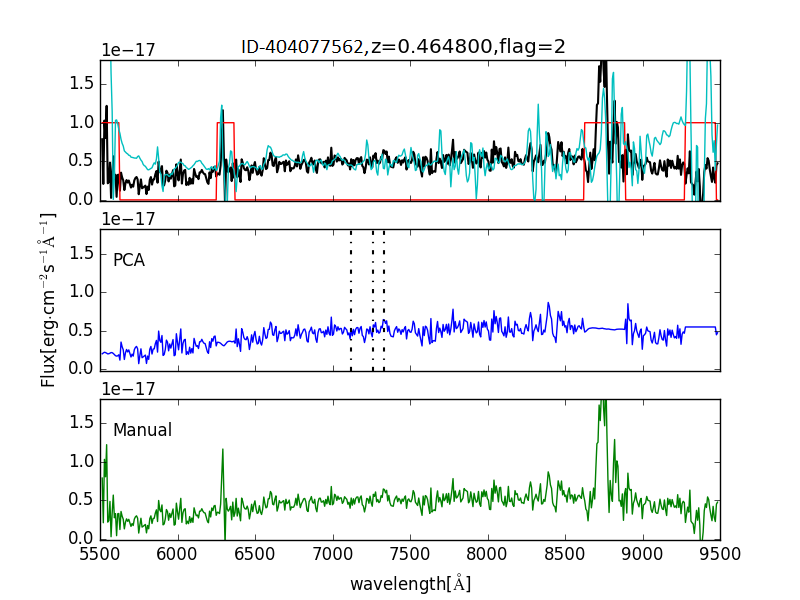}
\caption{\small{Cleaning of a Flag 4 (top) and a Flag2 (bottom) VIPERS spectrum: for each panel, the upper plot is the observed (sky subtracted) spectrum (thick black), superposed to the mask (red straight lines) and the rescaled sky residuals spectrum (thin cyan); the middle plot shows the automatic cleaning, with the expected position of the [OII], H$\beta$ and [OIII] lines marked in black by the dash-dotted lines; the bottom plot is the manually edited spectrum.}}\label{ex1}
\end{figure*}

Fig.~\ref{ex1} shows this comparison for two spectra of different quality. The black spectrum in the upper plot of Fig.~\ref{ex1}-top is a 
high signal-to-noise sky subtracted spectrum; the corresponding statistical PCA sky residuals spectrum  is overplotted in cyan; the red line shows the mask as computed by the cleaning procedure.   The resulting automatically masked and PCA repaired spectrum is shown in the middle plot of Fig.~\ref{ex1}-top (blue): within the regions of the mask, the observed spectrum has been replaced by the corresponding portion of the rest-frame PCA cleaned spectrum. The black dotted-dashed vertical lines highlight the wavelengths where some important emission lines are expected given the redshift. In these regions the repairing (if any) is not applied.  
The lower plot of Fig.~\ref{ex1}-top shows the manually edited spectrum. Overall, the automatically cleaned and repaired spectrum is very similar to the manually edited one. In the region of strong OH lines, the PCA cleaning looks more aggressive because the reconstructed portion of the spectrum is noise free while the human selectively edited the spurious features.

The bottom group of panels of Fig.~\ref{ex1} is like the previous, but shows a spectrum with lower signal to noise.
In this case, the automatic cleaning is more precise with respect to the manual cleaning, especially around the 6300\smash{\AA} sky line and the zero order spectrum at 8700\smash{\AA}.

\section{Conclusions}
We have presented a novel algorithm based upon principal component analysis to identify and repair spectral defects (such as those deriving from a non-perfect sky subtraction) in large sets of galaxy spectra.  We have implemented the procedure for the VIPERS dataset and tested its performance extensively against conventional manual spectral masking.  The data products produced by this work are part of the VIPERS second data release (PDR-2)  \citep{scodeggio16}.

The PCA algorithm characterizes a dataset with a compact set of components without specification of a model.  These components can represent the signal of interest but may also describe unwanted systematic effects as we explored in this work.   With the advent of spectroscopic surveys collecting millions of spectra, the use of automated procedures is becoming unavoidable to guarantee the efficient and accurate treatment of the data. 

\bibliographystyle{aa}
\bibliography{biblio_gg,biblio_VIPERS_v3}

\end{document}